\documentclass[twoside,twocolumn]{article}

\usepackage{graphicx}

\usepackage[sc]{mathpazo} % Use the Palatino font
\usepackage[T1]{fontenc} % Use 8-bit encoding that has 256 glyphs
\usepackage{microtype} % Slightly tweak font spacing for aesthetics

\usepackage[english]{babel} % Language hyphenation and typographical rules

\usepackage[hmarginratio=1:1,top=32mm,columnsep=20pt]{geometry} % Document margins
\usepackage[hang, small,labelfont=bf,up,textfont=it,up]{caption} % Custom captions under/above floats in tables or figures
\usepackage{booktabs} % Horizontal rules in tables

\usepackage{lettrine} % The lettrine is the first enlarged letter at the beginning of the text

\usepackage{enumitem} % Customized lists
\setlist[itemize]{noitemsep} % Make itemize lists more compact

\usepackage{abstract} % Allows abstract customization
\usepackage{titlesec} % Allows customization of titles
\renewcommand\thesection{\Roman{section}} % Roman numerals for the sections
\renewcommand\thesubsection{\roman{subsection}} % roman numerals for subsections
\titleformat{\section}[block]{\large\scshape\centering}{\thesection.}{1em}{} % Change the look of the section titles
\titleformat{\subsection}[block]{\large}{\thesubsection.}{1em}{} % Change the look of the section titles

\usepackage{fancyhdr} % Headers and footers
\usepackage{titling} % Customizing the title section

\pretitle{\begin{center}\Huge\bfseries} % Article title formatting
\posttitle{\end{center}} % Article title closing formatting
\title{Pickup and interference in particle astrophysics experiments: techniques and tools} % Article title
\author{%
\textsc{John McMillan} \\[1ex] % Your name
\normalsize Department of Physics and Astronomy, The University of
Sheffield\\ Sheffield, South Yorkshire, S3 7RH, Great Britain \\ % Your institution
\normalsize {j.e.mcmillan@sheffield.ac.uk} % Your email address
}
\date{\today} % Leave empty to omit a date
\renewcommand{\maketitlehookd}{%
\begin{abstract}
\noindent Contamination of signals by pickup and interference is a recurrent problem in physics experiments, the suppression of which consumes much time and effort.  Techniques are presented which allow the time-domain assessment of mitigation techniques. A pick-off circuit is presented which facilitates the
investigation of AC power-line borne transients without
danger to the experimenter or damage to oscilloscopes.
The circuit is simple to construct and low-cost. The output of this circuit may be used to veto or identify intereference produced events so that they are not subsequently included in analysis.
\end{abstract}
}

\begin{document}

% Print the title
\maketitle

\section{Introduction}

In the conduct of particle astrophysics experiments, the experimenter
must deal with pulsed signals from a range of detectors.
The contamination of these signals with interference and pickup
is a recurrent problem and the suppression of interference consumes
considerable time and effort.

Doubtless, this is also true in a variety of other disciplines. The major
difference being that particle astrophysics projects are often
undertaken in less than ideal experimental conditions, rather than in
the comfort of a purpose built laboratory, and the interference level
is more severe.  Consequently, the advice
given here will almost certainly find application in related disciplines.

It is not the intention of this note to provide a universal tutorial
on interference, pickup and techniques for its mitigation, as these can be found
elsewhere~\cite{Ba87a,Gr07a,Ot09a,Pa06a,Ta71a,Vi04a,Vi07a}.
Hemming~\cite{He92a}
provides useful information for those planning a new experiment or
facility.
Fitch~\cite{Fi76a} provides a particularly useful analysis of problems
with diagnostics in nanosecond pulsed power systems,
while Kerns~\cite{Ke92a} provides an anecdotal account of problems at an
accelerator laboratory.

Rather, this note is intended to present diagnostic techniques which allow
the quantification of interference, so that the experimenter can judge
whether mitigation techniques being applied are having the desired effect.
Additionally, there is the possibility of vetoing interference generated events, or
identifying them as such, so that they can be dealt with in
subsequent analysis.

The signal pulses typically have duration of a few nanoseconds to a
microsecond.  Some pulses are
simple yes/no trigger signals which need to be
counted or registered.  Others contain pulse height or shape
information which needs to be measured and is normally related
to energy deposited.  These signals are produced by semiconductors or
photomultipliers and can have amplitudes of only a few millivolts.

 In all cases it is likely
that these pulse signals must be monitored for a considerable period
of time, from hours up to some years in the worst cases.
It is in the nature of the subject that interest lies in rare events.
Hence, the
contamination of the recorded data with a significant proportion of
spurious events often leads to the failure of the experiment.

The interference can either be propagated by electromagnetic waves at radio
frequencies (RF) or it can be power-line borne.
More rarely it can be transmitted over signal or data circuits.
It can either be a continuous RF-signal or, more difficult to analyse,
be comprised of sporadic pulses.
However it is propagated, there will be a route by which it couples into
the detector output, either producing unwanted false events or
masking genuine signals.

In the majority of cases, the source of the interference is beyond the
control of the experimenter.  It may be generated by electrical
equipment in adjacent laboratories or buildings, by
load switching by the
utility supplier, or it may even be produced by lightning strikes.

For particle and nuclear physics applications, the most
problematic transients are of sub-microsecond duration, as these
closely mimic expected signals from detectors.  They correspond to
frequencies in the 10MHz -- 1GHz range, and in which broadcasts, mobile telephony and data services are operating.

\section{Power-line Filters}
It is impossible to overemphasise the importance of providing power-line filters.
The construction of an experiment or experimental
facility without substantial power-line filtering is simply a waste of
experimenters' time.  Commercial instrument crates and power supplies usually incorporate power-line filters but these are of variable quality and may be inadequate.  Low voltage supplies intended for powering computers or logic circuitry need only have sufficient filtering that the logic levels are not compromised.  When working with millivolt signals in a harsh environment, more substantial filters should be employed.

Power-line filters are industrially available
from a number of manufacturers.  They typically feature two or three
stages of low pass LC $\pi$-filter and can provide 40dB or more of
attenuation for signals in the 10MHz region.  The filter should be
conservatively rated to handle sufficient current for the experiment
or facility.

If there is a specific item of equipment which is known to be a source
of power-line noise and which is under the control of the experimenter, it
is worth providing a separate filter for the power-line leading into it.
This category includes pulsed particle generators, magnet
power supplies, pumps and almost anything with a high power motor.
The intention is that this will reduce interference
contaminating the supply to data acquisition systems.

Note that these filters go further than simple surge protectors.
These latter are semiconductor or similar devices intended to protect
equipment from over-voltage surges and which have non-ohmic response
such that above a specified voltage they dump current.  They are useful
for the function for which they are designed, and can be used
effectively to complement a filter, but they are no replacement.

Similarly, 'uninterruptible power supplies' (UPS) are just that.  They
are a system of rechargeable batteries and inverters intended to
provide continuous power to equipment in the event of power outages.
Such devices are no substitute for a filter and their rapid switching
during outages is likely to exacerbate interference problems.

\section{Antenna for Detecting RF}
To detect the presence of RF electromagnetic pulses, a simple whip
antenna is normally sufficient.  This can be 10 or 20cm of stiff wire
or a telescopic aerial of the type used on portable radios.  It should
be connected to the core of a length of 50$\Omega$ cable terminated at
the far end.  The output from this can be displayed on an
oscilloscope.  If there is considerable low-frequency signal present,
it is sometimes useful to crudely filter the antenna signal by
coupling it through a capacitor of about 10pF.

Typical laboratories will have continuous stream of pulses in the millivolt
region with larger pulses reaching into the volts or tens of volts.
Depending on the nature of the RF signals present, it may be possible to
diagnose their origin from the waveform.
It may also be possible to move the antenna and determine source
and directionality.

If the RF interference has significant continuous content of a
specific frequency, it might be worth using a dipole, a loop antenna or even a
directional Yagi antenna as a diagnostic tool.

The output from the antenna should be examined in the time domain on an oscilloscope, while comparing detector signals on other channels.  It can be quickly established at what level the detector signals are affected by RF interference.

\section{Power-line Pick-off Circuit}
Power-line noise presents a monitoring problem which makes diagnostics and
determining the success of mitigation attempts difficult.
Simply connecting the input of an oscilloscope
to the power-line supply is dangerous and likely to damage the instrument.
Using a high-voltage probe appears possible, but these are intended
for measuring high voltage signals rather than transients
superimposed on high voltage AC supplies.  They have large divider
ratios which reduce the transients down to unmeasurable levels.

A number of manufacturers offer power-line transient analysers.
These instruments are designed to record voltage trends,
dropouts, and power quality over a long period of time.
Typically, they provide a log with time stamps of
transients and outages which  can be analysed later.
They do not provide an
output that can be viewed in the time domain
with an oscilloscope, in order
to compare power-line transients with experimental
signals.

To address the monitoring problem, a simple pick-off
circuit was developed.
Being simple, the circuit presented is also of negligible cost
compared with commercial power-line transient analysers.

\subsection{Design and Construction}

The idea for the circuit is taken from Paul and Hardin
\cite{Pa88a}, who identify a method of isolating the measurement
from the power source, but who are interested in frequency
domain measurements of interference over a wide range.
The HF current probe presented by Wyatt~\cite{Wy12a} has
also been influential but is intended for a different application.
A simpler circuit, denoted ''garbage detector'' can be found in Hayes and Horowitz~\cite{Ha89a}. It does not use nanosecond pulse techniques and its operation will be dependent on the high frequency response of the power transformer used.

\begin{figure}
\begin{center}
 \includegraphics[width=\columnwidth]{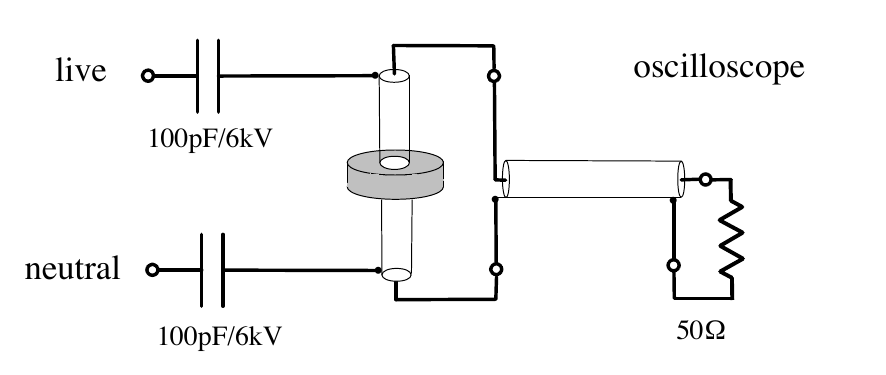}
\end{center}
 \caption{
 \label{schematic}
Schematic of pick-off circuit.
          } %end of caption
\end{figure}
The schematic of the pick-off
circuit can be seen in figure~\ref{schematic}.
It
is intended to pass signals in the 10MHz -- 1GHz range, while effectively
suppressing the power waveform at 50 or 60Hz and its immediate
harmonics.
The circuit comprises an isolating transmission line transformer
which is coupled to the power-line supply through low value high voltage
ceramic capacitors.

The transformer is the simplest possible
1:1 Guanella transmission line transformer \cite{Wi59a,Se01a}.
Alternatively,
if better impedance matching was of interest,
the isolation transformer proposed by Winningstad~\cite{Wi59a}
could be used, but this requires $Z_{0}/2$ transmission line which is
not readily available.  If lower frequency operation was required, a
design based on the isolation transformer proposed by Kaplan~\cite{Ka84a}
would be effective.

The transformer was made by winding five turns of RG-174 coaxial cable
through a toroidal ferrite core.
For those unfamiliar with nanosecond pulse transformers, it really is
arranged as drawn, with the 'primary circuit' formed by the coaxial braid
and the 'secondary circuit' formed by the inner core.
In the interests of simplicity of drawing, the schematic shows
only a single penetration of the core.

The toroid was a 23mm outside diameter
Ferroxcube TN23/14/7-4C65, which is a low permeability NiZn ferrite
intended for use in
wideband transformers. Similar materials are available from other
manufacturers.
\begin{figure}
\begin{center}
 \includegraphics[width=\columnwidth]{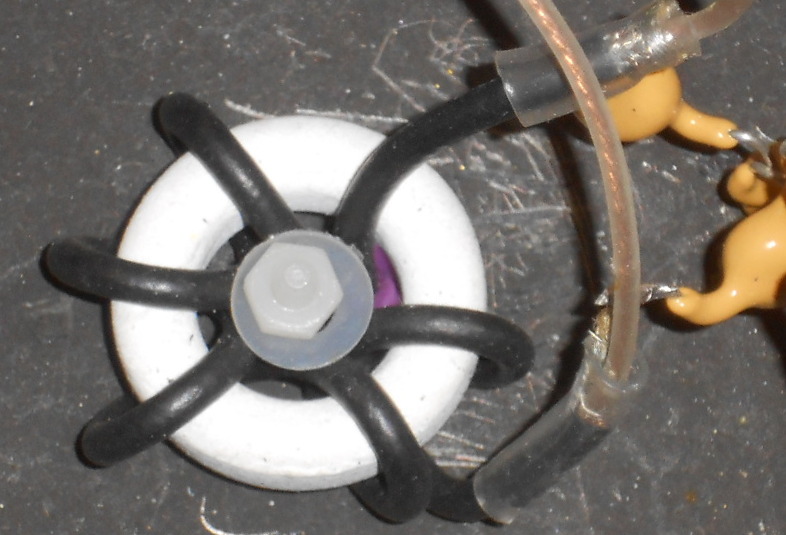}
\end{center}
 \caption{Transmission line transformer
 \label{photo}
          } %end of caption
\end{figure}
If very high voltage spikes are expected, the output of the transformer
could be further fitted with clamp diodes to prevent damage to the
oscilloscope.

A unit containing three of the circuits described above was
constructed in an insulated box.  The three circuits were wired
live--neutral, live--earth and neutral--earth. The connection to
the power-line was via an IEC 60320 C13 socket and the
three output connections
were BNC sockets.  By having three separate circuits, comparisons
can be made to determine whether transients are differential
live-to-neutral or common mode with respect to earth.
%
% Only introduce this if a referee requires it.
% otherwise stick with IEC.
%
%IEC terminology is Live, neutral, earth. North American practice
%Line, neutral ground.
%
\subsection{Response of the Pick-off Circuit}
The circuit presented does not provide a flat response to
transient signals at all frequencies.
In use, it is necessary to know how the signal observed is related
to the input.

A 1V pulse with risetime 0.7ns, fwhm 3.5ns and fall
4.0ns was applied to the input.  This produced an output pulse
of amplitude 280mV, risetime 1.0ns, fwhm 3.8ns and fall
2.2ns. It also had substantial ringing on its tail.

\begin{figure}[h]
\begin{center}
 \includegraphics[width=\columnwidth]{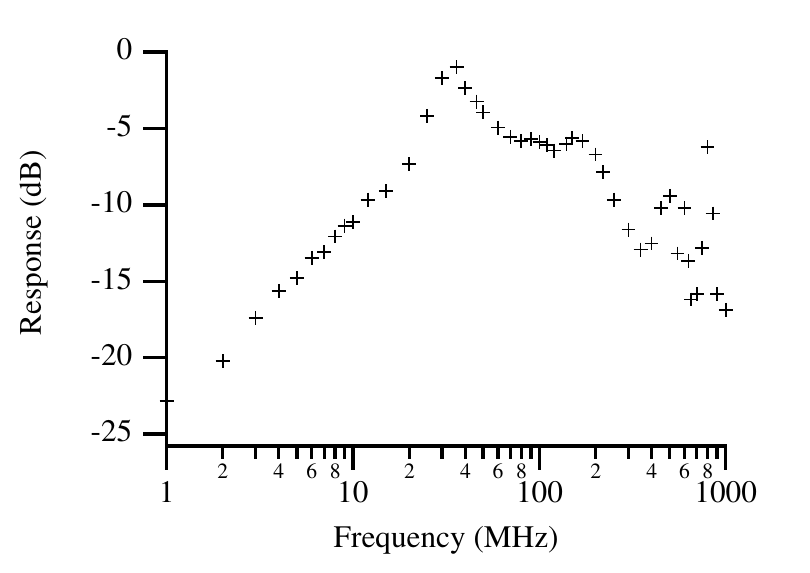}
\end{center}
 \caption{
 \label{response}
Measured response of pick-off circuit.
          } %end of caption
\end{figure}

Applying sine waves to the input and
measuring the input and output amplitudes gave the
frequency response function shown in figure~\ref{response}.
It is demonstrably far from flat. In the region above
600MHz there are clear
resonances which are due to the power-line input cable
being inadequately matched in its characteristic impedance.

By understanding the
response in the frequency domain and
observing transient behaviour in the time domain,
useful diagnostics of power-line transients can readily be obtained.

\section{Using the tools}

Either the antenna or the power-line pickoff circuit can be used to trigger an oscilloscope and other channels can be used to examine the detector signals. In this way vulnerabilities can be determined and any mitigation techniques applied can be assessed.

Further, either the antenna or the power-line pickoff circuit can be connected to a discriminator to generate
a logic flag, which can be used in event-based experiments to identify
those events which are
coincident with RF or power-line transients.
This logic flag can be used to veto interference generated events, or
it can be recorded by the data acquisition system so that these events can be identified later and excluded from
subsequent analysis.

 This approach is particularly useful when severe interference is encountered which is beyond the control of the experimenter.  Obviously it results in a reduction of the effective on-time of the experiment.
 The reduction may itself be modulated by some anthropogenic activity and care must be taken in the analysis of this.
\section{Acknowledgments}
I would like to thank all those who have contributed to this project over many years. I would also like to thank Steve Collins for useful discussions.
\bibliographystyle{unsrt}
\bibliography{pickup}

\end{document}